\documentclass[twocolumn,usenames,dvipsnames]{revtex4}

\usepackage{amsmath,amssymb,amsfonts} 
\usepackage{bm}
\usepackage[colorlinks=true, linkcolor=blue, urlcolor=blue, citecolor=blue]{hyperref}
\usepackage{graphicx}
\usepackage{color}
\usepackage{sidecap}
\usepackage{tikz}
\usepackage{cleveref}
\usepackage{lettrine}

\begin{document}

\title{Precise control of magnetic fields and optical polarization in a time-orbiting potential trap}
\author{A. J. Fallon}
\author{C. A. Sackett}
\email[Email: ]{cas8m@virginia.edu}
\affiliation{Department of Physics, University of Virginia, Charlottesville, VA 22904}

\date{\today}

\begin{abstract}
A time orbiting potential trap confines neutral atoms in a rotating magnetic field.
The rotation of the field can be useful for precision measurements, since it
can average out some systematic effects. However, the field is more difficult to 
characterize than a static field, and it makes light applied to the atoms have a
time-varying optical polarization relative to the quantization axis. These problems
can be overcome using stroboscopic techniques, where either a radio-frequency field or
a laser is applied in pulses that are synchronized to the rotating field. Using
these methods, the magnetic field can be characterized with a precision of
10~mG and light can be applied with a polarization error of $5\times 10^{-5}$.
\end{abstract}

\maketitle

\section{Introduction}

Magnetic traps are convenient tools for confining atoms over long time scales. The time-orbiting potential (TOP) 
technique offers the unique feature that the magnetic field at the center of the trap rotates rapidly in time \cite{Petrich95}.
Because the time average of their magnetic moment is zero, the trapped atoms have reduced sensitivity 
to low-frequency environmental fields. This makes TOP traps appealing for many types of precision
measurements. Indeed, bias field reversals are a standard
feature in measurements such as parity violation \cite{Guena05}, 
electric dipole moment searches \cite{Chupp19}, and other searches for new physics
\cite{Safronova18}. The TOP trap provides such reversals naturally and at a high frequency. It is not surprising,
then, that TOP traps have been considered for various types of precision measurement \cite{Crane01,Arnold04,Gupta05,jree2005}.

However, it is critical for many applications that the bias field be well characterized.
Ideally, the field should rotate in a well-defined plane with a constant magnitude and rotation rate.
If this is the case, then the time average of the atomic moment will be accurately zero, and it will be possible
to predict the instantaneous direction of the moment at any given time. Such characterization can be challenging,
because it is usually not possible to place a conventional magnetometer in situ at the location
of the atoms. In this paper, we present several methods to address this problem, and show that the magnetic
field can be optimized to an accuracy of about 10~mG. We also consider the problem
of applying light with a well-characterized polarization to the atoms, and show that even with a rotating bias field,
this can be achieved with an accuracy of about $5\times 10^{-5}$. 

Our own interest is in the application to tune-out wavelength spectroscopy \cite{Leblanc07,Arora11}. In this technique,
the trapped atoms are illuminated by a laser beam, and the laser is tuned to a frequency where the
ac electric polarizability of the atoms vanishes. By measuring this frequency precisely, information 
about the atomic matrix elements can be obtained. Our previous experiments \cite{rleo2015} took advantage of the 
rapid bias field reversal in a TOP trap in order to suppress the vector component of the polarizability,
which depends sensitively on the polarization of the applied light. For a new experiment \cite{afal2016}, we wish
to measure the vector polarizability, so it is necessary to characterize both the trap field and the
light polarization.  Although the discussion is centered on our particular requirements, we believe that the methods 
presented will be useful for many types of experiments, since it is often necessary to apply well-characterized
light to atoms in a well-known magnetic field. 

The paper is organized as follows: Section II describes the
TOP trap and characterizes how imperfections can distort the bias field. We present a radio-frequency
spectroscopy technique that can be used to characterize the bias field as a function of time,
and another technique that uses the motion of atoms in the trap to measure
an additional component of the ambient field. Section III describes
a method to apply polarized light to the trapped atoms, accounting for imperfection
from both the light polarization and the magnetic field direction.
Finally, Section VI provides a summary, conclusions, and outlook for further
improvement.

\section{Magnetic field characterization and control}

	Our apparatus uses a modified TOP configuration. Its implementation and operation have been described previously \cite{jree2005,Burke10}, but we summarize relevant details here. The basic trap is formed by a rotating bias field $\textbf B_0$ and an oscillating linear quadrupole $\textbf B_1$. These support the atoms against gravity in the vertical $z$ direction, and they provide approximately harmonic confinement in $z$ and the transverse direction $x$. An additional spherical quadrupole $\textbf B_2$ is applied which oscillates at a different frequency and provides adjustable weak confinement in the longitudinal direction $y$. Altogether, these fields can be expressed as 
$		\textbf B_{tot} = \textbf B_{0} + \textbf B_{1} + \textbf B_{2} $
with
	\begin{equation} \label{eqn:b0}
		\textbf B_{0} = B_{0} \big( \sin\Omega_{1} t \hat x + \cos\Omega_{1} t \hat z \big)
	\end{equation}
	\begin{equation} \label{eqn:blin}
		\textbf B_{1} = B'_{1} \big( z \hat z - x \hat x \big) \cos\Omega_{1} t
	\end{equation}
	\begin{equation}
		\textbf B_{2} = B'_{2} \big( 2y \hat y - x \hat x - z \hat z \big) \cos\Omega_{2} t.
	\end{equation}
The trapping potential is given by $\mu \langle |\textbf B_{tot}| \rangle$, where
the angle brackets denote a time average and $\mu$ is the magnetic moment of the spin state. 
Our experiments use the $F=2, m_F =2$ hyperfine state of $^{87}$Rb, so that
$\mu$ is approximately equal to the Bohr magneton $\mu_B$. In general the time average must
be calculated numerically, but if the atoms remain close to the origin then it is accurate
to Taylor expand $|\textbf B_{tot}|$ to second order and perform the
time average analytically. The result is 	
	\begin{align} \label{eqn:bmag}
		\langle | \textbf B | \rangle \approx B_{0} & - \frac{1}{2} B'_{1} z + \left ( \frac{3B_{1}^{\prime \hspace{2pt} 2}}{16B_{0}} + \frac{B_{2}^{\prime \hspace{2pt} 2}}{4B_{0}} \right ) x^{2} \nonumber \\
		& +  \frac{B_{2}^{\prime \hspace{2pt} 2}}{B_{0}} y^{2} + \left ( \frac{B_{1}^{\prime \hspace{2pt} 2}}{16B_{0}} + \frac{B_{2}^{\prime \hspace{2pt} 2}}{4B_{0}} \right ) z^{2}.
	\end{align}
Here we assume that $\Omega_1$ and $\Omega_2$ are approximately incommensurate, so that no cross terms survive the time average. Experimentally we use $\Omega_1 = 2\pi\times 12.8$~kHz
and $\Omega_2 = 2\pi\times 1$~kHz.  Typically we use
$B_0 \approx 24$~G and we set $B_1' \approx 30.7$~G/cm such that the linear term in the TOP potential cancels the gravitational potential $mgz$. We set $B_2' \approx 2.5$~G/cm to 
provide an oscillation frequency $\omega_y \approx 2\pi\times 1$~Hz.  The measured $\omega_x$ and $\omega_z$ confinement frequencies are then
approximately $2\pi\times 5.1$~Hz and $2\pi\times 3.3$~Hz, respectively. In comparison,
Eq.~\eqref{eqn:bmag} predicts values of 4.9 and 2.9~Hz. The difference is due to  non-uniformity of the bias field $B_0$, but this
has negligible impact on the work discussed here since it alters the spatial variations of the field but not the field itself at the potential minimum.

A number of other non-idealities do impact the field experienced by the atoms. The rotating bias field components are produced by two separate coils. These
coils may not be perfectly orthogonal, their fields may have different amplitudes, 
and their phase difference may differ from $\pi/2$. In addition a dc background field
may be present.
All of these effects can introduce time-dependent variations in the field magnitude and
direction at the position of the atoms. The goal here is 
to characterize and control these effects. 

The two components of the bias field are produced by long rectangular coils oriented near $\pm 45^\circ$ from vertical. We express these components as
\begin{align}
    \textbf B_{0a} = \frac{B_{0}}{\sqrt{2}} & (1+\Delta)\big[(1+\psi_1) \hat x - (1-\psi_1)\hat z \big] \nonumber \\
    & \times \sin\left( \Omega_1 t - \frac{\pi}{4} + \xi_1 \right)
\end{align}
and
\begin{align}
    \textbf B_{0b} = \frac{B_{0}}{\sqrt{2}} & (1-\Delta) \big[ (1-\psi_2) \hat x + (1+\psi_2)\hat z \big] \nonumber \\
    & \times \sin\left( \Omega_1 t + \frac{\pi}{4} + \xi_2 \right),
\end{align}
where $\Delta$ characterizes the amplitude mismatch, the $\psi_i$ are small 
angular deviations from the ideal orientation, and the $\xi_i$ are phase offsets from 
the $\textbf B_1$ quadrupole oscillation. 
It is useful to define common and differential variables
$\psi = (\psi_1 + \psi_2)/2$, $\psi' = (\psi_1-\psi_2)/2$,
$\xi = (\xi_1+\xi_2)/2$, and $\xi' = (\xi_1-\xi_2)/2$.

Expanding to first order in these small variables, we find a total bias field of
    \begin{align}
	    \frac{\textbf B_{0}}{B_0} =  & \Big[ \big( 1 + \xi' + \psi' \big) \sin\Omega_1 t - \big( \Delta - \xi + \psi \big) \cos\Omega_1 t \Big] \hat x \nonumber \\
	    & + \Big[ \big( 1 - \xi' - \psi' \big) \cos\Omega_1 t - \big( \Delta + \xi - \psi \big) \sin\Omega_1 t \Big] \hat z.
    \end{align}
To this we add the $\textbf B_1$ quadrupole field from \eqref{eqn:blin} 
and an environmental field 
$\textbf B_E = B_{Ex}\hat x + B_{Ey} \hat y + B_{Ez} \hat z$ with 
$|B_{Ei}| \ll B_0$. 
We then calculate the TOP potential using the same time-averaging procedure as
before. We omit the $B_2$ field since it is an order of
magnitude smaller than the $B_1$ quadrupole. The result is 
\begin{align} \label{eq:timeavg}
    \langle | \textbf B_{tot} | \rangle = B_{0} \bigg\{  1 &  + \frac{1}{4}  \big( \Delta - 2 \xi + 2 \psi \big) qx + \frac{1}{4} \big( 2  - \xi' - \psi' \big) qz \nonumber \\
    & + \frac{3}{16} q^{2} x^{2} + \frac{1}{16} q^{2} z^{2} \bigg\},
\end{align}
with $q \equiv B_1'/B_0$. Here we keep terms to first order
in $\Delta, \psi, \psi', \xi$, $\xi'$, and $B_{Ei}/B_0$, except in the
$x^2$ and $z^2$ terms where the non-idealities are omitted.

The atoms will be trapped at the minimum of the total potential. Along $x$ the minimum
can be found directly as
    \begin{equation}
	    x_0 = -\frac{2 \left( \Delta - 2 \xi + 2 \psi \right)}{3 q}.
    \end{equation}
We take the vertical position $z_0$ as an independent
parameter. We can then express the time-dependent field magnitude at the center, to first order in non-idealities, as
	\begin{align} \label{eqn:rfspec}
	    | \textbf B_{tot} |(t) = B_{0} \bigg\{ 1 & + \frac{1}{2} qz_0 + q_{Ex} \sin\Omega_1 t + q_{Ez} \cos\Omega_1 t \nonumber \\
	    & +  \frac{1}{2} \big( qz_0 - 2 \xi' - 2 \psi' \big) \cos2\Omega_1 t \nonumber \\
	    & - \frac{2}{3} \big(  \Delta + \xi - \psi \big) \sin2\Omega_1 t  \bigg\}
    \end{align}
with $q_{Ei} = B_{Ei}/B_0$.
We see that the non-idealities combine to give oscillating contributions to $|\textbf B|$
that have different frequencies and phases. Measuring these different components therefore
provides information about the non-idealities, which can then be compensated with the goal
of producing a bias field that varies as little as possible. We see that it is not necessary
for all the non-ideal parameters to be zero, since the combinations $qz_0-2\xi'-2\psi'$
and $\Delta+\xi-\psi$ appear together. As long as the parameters are adjusted
to make $|\textbf B_{tot}|$ constant in time, the net bias field will rotate uniformly
as 
\begin{equation}
    \textbf B_{tot} = B_0\left(1+\frac{qz_0}{2}\right)
    \big(\hat x \, \sin\Omega_1 t + \hat z \, \cos\Omega_1t\big) + B_{Ey} \hat y.
\end{equation}
The amplitude shift due to $z_0$ is typically unimportant, so we do not attempt to 
measure or compensate for it.

	\begin{figure}
        \includegraphics{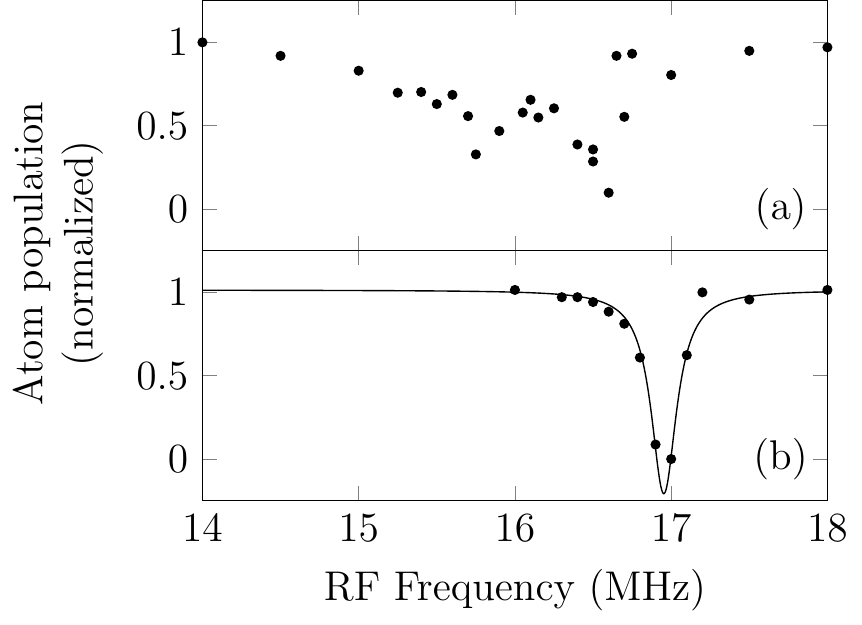}
		\caption{Radio-frequency spectra of trapped condensate atoms. The vertical axis shows the fraction of atoms remaining in the trap after rf is applied at the indicated frequency. (a) Spectrum obtained using a single long pulse of duration 200 ms. (b) Spectrum obtained
		using a train of 250 pulses each with 10~$\mu$s duration. The pulses are synchronized to the 12.8~kHz bias rotation frequency, so that the magnetic field has the same value during each pulse. At the delay time shown, the field magnitude happens to take on nearly its largest value. The curve is a Lorentzian fit.
		} 
		\label{fig:continuousrf}
	\end{figure}
	
Information about the magnetic field at the location of the atoms can be obtained
by driving the Zeeman transition $m_F = 2 \rightarrow m_F =1$ using a radio-frequency
field. Atoms making the transition are no longer supported against gravity and fall out of the trap. If the atoms form a Bose--Einstein condensate, the thermal broadening of the rf
spectrum will be negligible and the character of the spectrum will be determined entirely
by the variations in the magnetic field at the trap potential minimum.
Figure~\ref{fig:continuousrf}(a) shows the spectrum observed when a continuous
rf pulse is applied to an unoptimized trap. The broad and complicated lineshape 
indicates that the atoms experience considerable variations in the trap field,
making the resonant frequency vary over the course of the TOP period.

More detailed information can be obtained by applying a pulsed rf field, with the pulses
synchronized to the $\Omega_1$ trap frequency. In this way we obtain a snap-shot of the field value at a particular point in the cycle, using the same principle as the stroboscope. 
Figure~\ref{fig:continuousrf}(b) shows the spectrum obtained with a 10~$\mu$s pulse duration 
at a fixed delay with respect to the 80~$\mu$s oscillation period. The spectrum is much
narrower, with a width close to the 60~kHz transform limit of the pulse. The 
frequency at which the peak occurs indicates the instantaneous value of the field at that time.

\begin{figure}
		\includegraphics{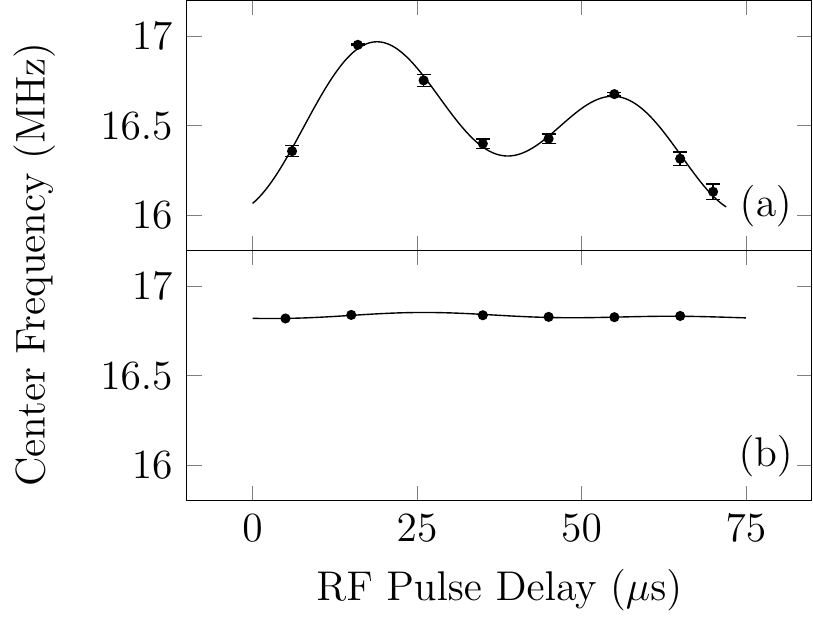}
		\caption{Time-dependence of the magnetic field magnitude during TOP field oscillation, as measured by the center frequency of spectra such as in Fig.~\protect\ref{fig:continuousrf}(b). Error bars are one-$\sigma$ errors from the
		fit. Solid curves are fits to the form of Eq.~\protect\eqref{eqn:rfspec}.(a) Initial variation in a trap using nominal driver current amplitudes.  (b) Variation after adjusting the oscillating terms in \protect\eqref{eqn:rfspec} to be zero. The residual oscillation corresponds to field variations of less than 10~mG.}
		\label{fig:bias}
	\end{figure}

To map out the field amplitude as a function of time, we take a series of spectra such as 
Fig.~\ref{fig:continuousrf}(b) with different time delays between the trap current oscillation and the rf pulses. A typical result is shown in Fig.~\ref{fig:bias}(a). We fit such data to a function 
with the form of Eq.~\eqref{eqn:rfspec}, where the amplitudes of each term are fit parameters.
The solid line in the figure shows the result, which generally fits the data well.

The fitted coefficients indicate how the parameters $B_{Ex}$, $B_{Ez}$, $q$ and $\Delta$ can
be adjusted to make $|\textbf B|$ constant in time. 
We do not adjust the $\psi_i$ or $\xi_i$ variables.
The environmental fields are controlled
using a set of bias coils, while $q$ and $\Delta$ are set by the quadrupole and bias current
amplitudes respectively. Figure~\ref{fig:bias}(b) shows a spectral measurement
of the field variations after the oscillating components have been minimized, 
showing that the transition frequency remains nearly constant during the bias rotation. 
Our measurement resolution 
is 5~mG, and we are able to zero each frequency component to
that level. This corresponds to a total rms field variation of about 10~mG.
	
 The rf spectroscopy technique is insensitive to the $B_{Ey}$ component,
 since it makes only a dc contribution to the field magnitude. However, we want to ensure
 that the field rotates in the $xz$ plane, so it is necessary to determine and null out the
 $B_{Ey}$ field as well. A way to achieve this is by applying a dc spherical quadrupole field
 	\begin{equation} \label{eqn:dcquad}
	    \textbf B_{Q} = B'_{Q} ( 2z \hat z - x \hat x - y \hat y)
	\end{equation}
to the atoms in the TOP trap. We focus on the resulting confinement potential along 
the $y$ direction, taking $x = z = 0$. Calculation of the time-averaged field magnitude as in 
Eq.~\eqref{eq:timeavg} yields
 	\begin{equation}
		\langle | \textbf B | \rangle = B_0 + \frac{B_2'^2}{B_0}y^2
		+\frac{B_Q'^2}{2B_0} \big(y-y_E\big)^2,
	\end{equation}
where $y_E \equiv B_{Ey}/B_Q'$ is the position where the net dc field is zero.
The minimum of the resulting potential occurs at position
	\begin{equation} \label{eqn:overshoot}
		y_{0} = y_E \frac{B_Q'^2}{B_Q'^2 + 2B_2'^2} = \frac{B_{Ey} B_Q'}{B_Q'^2 + 2B_2'^2}.
	\end{equation}

\begin{figure}
		\includegraphics{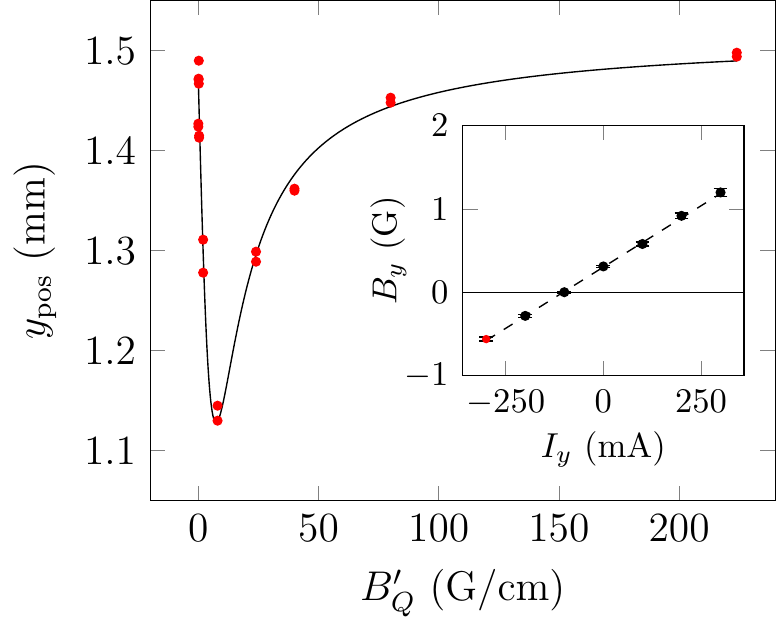}
		\caption{Using trap motion to determine the $B_{Ey}$ background field. In the main graph, data points are the observed trap positions as the dc
		gradient $B_Q'$ is slowly varied. The curve is a fit to 
		Eq.~\protect\eqref{eqn:overshoot} yielding $B_{Ey} = 0.56(2)$~G. The inset shows measured $B_{Ey}$ 
		values as a function of current $I_y$ through a pair of dc bias coils. The red point corresponds to the data in the main graph.}
		\label{fig:ybias}
	\end{figure}

    To find $ B_{Ey} $, we measure the condensate's position while varying $ B_{Q}' $ and fit the results to Eq.~\eqref{eqn:overshoot}. Typical data are shown in 
    Fig.~\ref{fig:ybias}(a), along with the fit curve. 
    Figure~\ref{fig:ybias}(b) shows the values of $B_{Ey}$ 
    obtained from the fit as current through a dc bias coil is varied. The slope of 
    the curve is consistent with the bias coil geometry, and the intercept allows us 
    to determine where $B_{Ey}$ is zero to an accuracy of 7~mG. We used a similar
    technique observing motion along the $z$ direction, and verified that the trap motion and rf spectroscopy techniques give consistent results for the $B_{Ez}$ component.

    The background magnetic fields and rotating bias field show good stability over long timescales without the need for regular adjustments. We observed drifts of less than 10 mG over several months of operation. However, the linear quadrupole amplitude $ \textbf B_{1}' $ does drift by about $30$~mG/cm over the course of days, making regular adjustments necessary. It is easy to see when $B_1'$ has shifted, because the $z$ position of the atoms changes. 

\vspace{0.5in}

\section{Optical polarization characterization and control}

\newcommand{\EE}{\mathcal{E}}

In addition to having a well-controlled magnetic field, we need to apply 
a light field with a well known and stable polarization. This is a critical
element for our tune-out wavelength studies, and it is important for other
precision measurements as well. For our experiments, we need to apply
$\sigma_+$ polarized light to the atoms with a polarization accuracy better than $10^{-4}$.

Two factors make polarization control challenging here. The first is that the bias
field at the atoms is rotating, so relative to the quantization axis the
light polarization is constantly changing. This can be addressed using the same
technique described above for rf spectroscopy, by applying short pulses of light that are 
synchronous with the magnetic field oscillations. If light polarization 
$\hat{\EE}$ is applied to the atoms, the polarization fidelity can be defined
as $F = \langle |\hat{\EE}^*\cdot \hat{\sigma}_+|^2\rangle$, 
where the angle brackets denote a time average over the direction of the field.
We use circularly polarized light travelling along $z$, with
$\hat{\EE} = (\hat{x}-i\hat{y})/\sqrt{2}$. The direction of the trap field
determines the $\hat{\sigma}_+$ vector
as $(\hat{x}'-i\hat{y})/\sqrt{2}$, for $\hat{x}' = \cos\Omega_1 t \hat{x}
+\sin\Omega_1 t \hat{z}$. If the light applied for time $\tau \ll 1/\Omega_1$, centered
on $t = 0$, then the time-averaged fidelity is
\begin{equation}
    F = 1- \frac{1}{48}\Omega_1^2 \tau^2.
\end{equation}
For $\Omega_1 = 2\pi\times 12.8$~kHz, this gives a negligible 
polarization error of $2\times 10^{-6}$ at a pulse duration of 120~ns.

The second challenging factor is that
optical polarizing elements are not ideal, so the light polarization reaching
the atoms will not be perfect. For instance, stress-induced birefringence of the vacuum window introduces polarization errors that are difficult to determine in situ
\cite{schott19}.
Similarly, waveplate retardances are not exact and can vary with 
temperature and light wavelength.

The behavior of the polarization can be characterized using 
the Stokes vector $[S_1, S_2, S_3]$, which can be related to the left-
and right-circular polarized electric field components $\EE_\ell$ and $\EE_r$ by
$S_1 = 2 \,\text{Re}(\EE_r \EE_\ell^*)$, 
$S_2 = -2\, \text{Im}(E_r E_\ell^*)$ and $S_3 = |\EE_r|^2 - |\EE_\ell|^2$. The $S_0$ 
Stokes parameter is here taken to be unity, 
and we normalize $S_1^2+S_2^2+S_3^2 = |\EE_r|^2 + |\EE_\ell|^2 = 1$. 
When the laser beam passes through a birefringent
element with retardance $\delta$ and axis at angle $\alpha$, the effect on $\textbf S$
is given by the Mueller matrix \cite{Chipman10}
\begin{widetext}
	\begin{gather}
	    M(\alpha,\delta) = \begin{bmatrix}  \cos^{2}2\alpha+\sin^{2}2\alpha\cos\delta & \cos2\alpha\sin2\alpha\left(1-\cos\delta\right) & \sin2\alpha\sin\delta \\  \cos2\alpha\sin2\alpha\left(1-\cos\delta\right) & \cos^{2}2\alpha\cos\delta+\sin^{2}2\alpha & -\cos2\alpha\sin\delta \\  -\sin2\alpha\sin\delta & \cos2\alpha\sin\delta & \cos\delta \end{bmatrix},
	\end{gather}
\end{widetext}
such that input $\textbf S$ is transformed to  $\textbf S' = M\textbf S$. 
The fidelity of the output polarization with respect to the initial state is given by
\begin{equation}
    F = \frac{1}{2}\left(1+\textbf S'\cdot \textbf S \right).
\end{equation}
In the case of weak birefringence $\delta \ll 1$, the fidelity can be calculated
to second order as
\begin{equation}
    F \approx 1- \frac{\delta^2}{4}\big[(S_1\sin 2\alpha - S_2 \cos 2\alpha)^2 + S_3^2\big],
\end{equation}
The error is zero for linearly polarized light aligned to the axis of the retarder,
but in general the fidelity decreases by a factor of order $\delta^2$. A similar error occurs for light passing through a waveplate if $\delta$ is interpreted as the birefringence error
and $\textbf S$ is the ideal output polarization. 
We observe typical values of $\delta$ to be $5\times 10^{-2}$ or greater,
which imposes a polarization error on the order of $10^{-3}$. 
It is therefore necessary to correct for these errors. 

 We prepare
the polarization state starting with linear polarization produced by a Glan--Taylor
polarizer, with an estimated error below $10^{-5}$ \cite{Takubo98}. The conversion to 
circular polarization
is achieved using a Fresnel rhomb, which is the most stable
retarder readily available \cite{Bennett70}. 
Using BK7 glass, the calculated wavelength variation of
the retardance is below $10^{-8}$~rad/nm, and the calculated temperature dependence is
about $4\times 10^{-6}$~rad/K. We verified experimentally that the retardance of
the rhomb is stable at our measurement sensitivity of $10^{-5}$. 

The retardance of the rhomb is not easily adjustable, so prior to the rhomb
we pass the light
through two Meadowlark Optics zero-order polymer 
retarders, one a quarter-wave plate and the other a half-wave plate. Both plates
are aligned with their axes close to the incident polarization axis, which limits
the sensitivity to retardance errors or drifts.  The polarization state exiting the
rhomb is then
\begin{equation}
    \textbf S_\text{rhomb} = 
    \begin{bmatrix} -2\alpha_1 \\ 4\alpha_2-2\alpha_1 \\ 1 \end{bmatrix}
    + O(\alpha^2)
\end{equation}
where $\alpha_1$ is the angle of the quarter-wave plate and $\alpha_2$ the angle of the 
half-wave plate. Any inaccuracies of the rhomb or polarization shifts from subsequent 
optical elements will give additional small contributions to $S_1$ and $S_2$.
We see, however that the two waveplate angles 
provide sufficient degrees of freedom to compensate for any such contributions,
allowing $S_1$ and $S_2$ to be tuned to zero.

It is useful to calculate the 
projection of the light polarization onto the atomic $\sigma_-$ and $\pi$ components,
in terms of the Stokes parameters and the relative orientation
between the laser beam and the magnetic field. The results are
\begin{align}
    |\EE_\pi|^2 & = \frac{1}{2}\left(1 + S_1\cos 2\phi + S_2 \sin 2\phi\right)\sin^2\theta \approx \frac{\theta^2}{2} \\
    |\EE_-|^2 & = \frac{1}{2}(1\!-\!S_3\cos\theta)\! -\! \frac{1}{4}\left(1\! +\! S_1\cos 2\phi \!+\! S_2 \sin 2\phi\right) \sin^2\!\theta \nonumber \\
    & \approx \frac{S_1^2+S_2^2}{4},
\end{align}
where the laser beam propagates at polar angles $(\theta,\phi)$ with respect to the 
field. We see that the $\pi$ polarization component depends primarily on alignment,
while the $\sigma_-$ term is set by the polarization optics. The polarization error
$1-F$ can be expressed here as $|\EE_\pi|^2 + |\EE_-|^2$.
	
This analysis shows that
in order to apply pure $\sigma_+$ light to the atoms, several conditions must be met.
First, the laser beam should be aligned to the $z$ direction of the trap.
Second, the laser pulse timing must be set so that the pulse center arrives when
the trap field points along $z$. Finally, the waveplate angles $\alpha_1$
and $\alpha_2$ must be adjusted to compensate for the birefringence of the vacuum
window and any other polarization errors. 

\begin{figure}
    \includegraphics[width=3.2in]{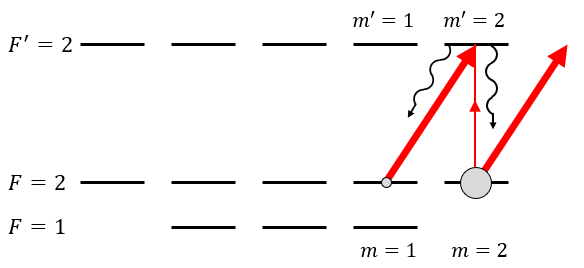}
    \caption{\label{fig:levels}
    Level diagram for polarization testing. Atoms are trapped in the $F=2, m = 2$
    state, where they cannot scatter $\sigma_+$ polarized light. Any contamination by
    $\pi$ or $\sigma_-$ light does lead to scattering and loss from the trap; the diagram shows $\pi$ light for illustration. Because of the scattering, a small population can be temporarily established in the $F = 2, m=1$ state, where the strong excitation
    to $m' = 2$ can destructively interfere with the excitation amplitude from $m = 2$. This leads to a suppression of scattering at high optical intensity.}
\end{figure}

In order to set these values precisely, we require a means to characterize the
polarization at the location of the atoms. As shown in Fig.~\ref{fig:levels},
our $^{87}$Rb atoms are trapped
in the $F = 2, m_F = 2$ ground state, and we measure the polarization fidelity by
tuning the laser to the $5P_{1/2}$ $F = 2$ level. This level has no state with
angular momentum projection $m = 3$, so pure $\sigma_+$ light does not
scatter from the atoms. We can then use the scattering rate as a measure of polarization
error, which is very sensitive since scattering 
even a single photon causes an atom to 
be removed from a Bose--Einstein condensate. 

\begin{figure}
        \includegraphics{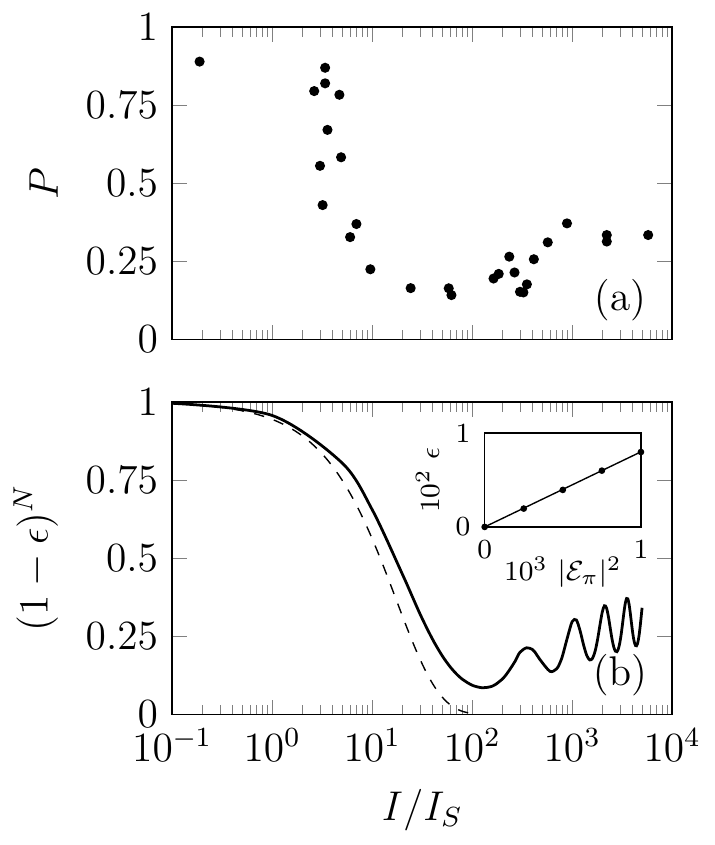}
		\caption{(a) Experimental measurements of atom loss. Data points show the fraction of atoms $P$ remaining in the trap after 1280 pulses of laser light at the indicated total intensity, relative to the saturation intensity $I_S$. (b) Numerical calculation of the survival probability after $N$ pulses $(1-\epsilon)^N$ for $N=1280$. The solid curve shows the result from the optical Bloch equations for a
		polarization impurity $\EE_\pi = 2\times10^{-4}$. The dashed curve shows the behavior that would be
		expected in the absence of dark-state formation. The inset shows that the loss
		$\epsilon$ depends linearly on the polarization impurity, here calculated at $I = 100 I_S$. 
		The slope $d\epsilon/d|\EE_\pi|^2$ is approximately 9.5 at the first minimum.}
		\label{fig:model}
	\end{figure}

To make the measurement, we apply up to 4000 light pulses, each of duration 120 ns
and with a period of $2\pi/\Omega_1$.
We then measure the fraction of atoms remaining in the trap.
We observe the scattering rate for near-$\sigma_+$ light to be a complicated 
function of the total intensity, as seen in Fig.~\ref{fig:model}(a). This
is due to the formation of a dark state. For example, Fig.~\ref{fig:levels} shows a case where 
a small amount of $\pi$ light
is present. This excites atoms into the $m' = 2$ state, where they can decay to the 
$m = 1$ ground state and eventually fall out of the trap. 
However, the atoms do not move significantly during the short laser pulse, 
so atoms with $m = 1$ undergo a strong excitation to the 
$m' = 2$ excited state from the $\sigma_+$ light. For the proper spin superposition 
$|\psi\rangle = \sum c_i|m_i\rangle$, the excitation amplitude
from $m = 1$ to $m' = 2$ can cancel the amplitude from $m = 2$ to $m' = 2$, leaving
the state $|\psi\rangle$ dark. 

The trapped atoms experience a Zeeman splitting of about 17~MHz, as seen in 
Fig.~\ref{fig:bias}. This causes the phases in $|\psi\rangle$ to change in time,
so in order to maintain the 
dark state it is necessary for the optical Rabi frequency of the light 
to be comparable to the Zeeman splitting.
This corresponds to an intensity $I$ of roughly ten times the saturation intensity $I_S$, 
which agrees with the 
measured intensity where the atom loss starts to level out. The Zeeman shift 
causes substantial dephasing during the 80~$\mu$s between laser pulses, and measurements
confirm that that each pulse has an independent effect on the atoms.

We have analyzed the formation of the dark state by solving the optical Bloch equations for
the thirteen relevant atomic states involved \cite{Foot05}. This includes the $F = 2$ ground states,
the $F' = 2$ excited states, and the $F=1$ ground states which can be populated by
spontaneous emission. We model the evolution during a single pulse of the light, and
determine the fraction of atoms $\epsilon$ lost from the initial $m =2$ state as a function of the
intensity components $I_i$, with $I_\pi, I_- \ll I_+$. Figure~\ref{fig:model}(b) shows
how the loss depends on the total intensity, and the shape of the curve agrees reasonably well with
the experimental observations. We do not clearly observe the predicted 
oscillations at high intensity, but
it is likely they are washed out by experimental intensity noise. 
The inset shows that the loss $\epsilon$ depends linearly on 
the polarization impurity.

The formation of the dark state limits the sensitivity of our polarization measurement,
since we cannot arbitrarily increase the laser intensity without saturating the loss rate. 
Instead we experimentally adjust the intensity to locate the value where the loss rate is
largest, and then use the Bloch equation model to determine the polarization impurity
corresponding to the measured loss. This calibration depends differently on the $\pi$ and
$\sigma_-$ components, with the loss rate always being greater for $\sigma_-$ light.
For both polarizations, the loss rate
maximum occurs at $I \approx 100 I_S$. At that minimum we evaluate the loss per pulse as
$\epsilon = \kappa_i |\EE_i|^2$, finding $\kappa_\pi \approx 9.5$ and $\kappa_- \approx 18$.
To be conservative, we assume that the impurity is all $\pi$ light to set an upper bound. 
We are then able relate the
measured atom survival probability $P = (1-\epsilon)^N$ 
to the polarization impurity $|\EE_\pi|^2$ via
\begin{equation}
    |\EE_\pi|^2 = \frac{1-P^{1/N}}{\kappa_\pi},
\end{equation}
for number of pulses $N$.
For the data of Fig.~\ref{fig:model}(a), we obtain $|\EE_\pi|^2 \approx 1.5\times 10^{-4}$.

\begin{figure}
        \includegraphics{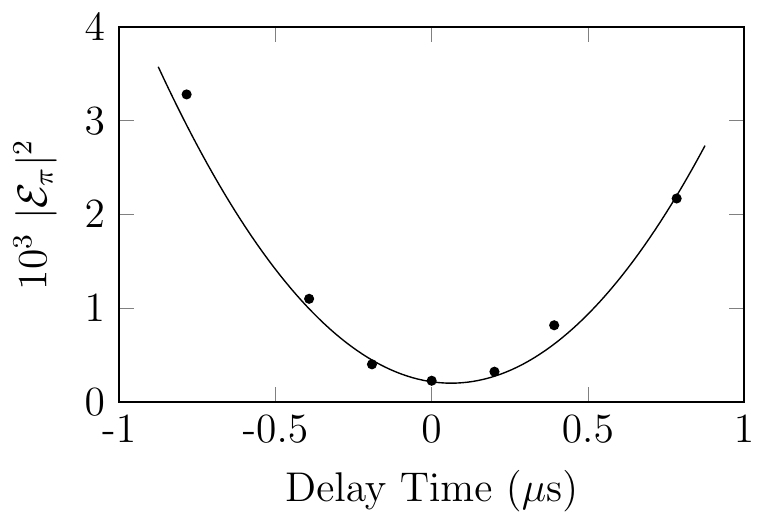}
		\caption{Dependence of polarization error on beam alignment. Points show the fraction of $\pi$ polarized light at the atoms,
		determined as described in the text. The angle $\theta$ between the laser beam
		and the rotating field is varied by adjusting the time $t$ at which the light pulse is centered. The curve is a parabolic fit giving $|\EE_\pi|^2 = 0.61(3)\cdot \Omega_1^2 t^2$, 
		in reasonable agreement with the expectation $|\EE_\pi|^2 = \theta^2/2$.}
		\label{fig:alignment}
	\end{figure}

Following this procedure, we can optimize the light polarization, pulse timing, 
and beam direction to minimize the polarization error. 
For example, Fig.~\ref{fig:alignment} 
shows how the atom loss varies when the delay time of the light pulse is changed.
This corresponds to varying the angle between the beam and the 
rotating field, with $\Delta\theta = \Omega_1 \Delta t$. The polarization error
varies like $\theta^2/2$, as expected. The optimum delay time corresponds to the minimum
of the curve.
After optimizing all parameters in this way, we consistently obtain a loss rate corresponding to 
$|\EE_\pi|^2 = 5\times 10^{-5}$. 
Alternatively, if we assume the polarization impurity to be
$\sigma_-$, we infer $|\EE_-|^2 = 3\times 10^{-5}$. 

To confirm this result, we reversed the handedness of the light 
by rotating the initial polarizer by $90^\circ$, and offset
the pulse timing by a half-period $\pi/\Omega_1$. 
We then re-optimized the waveplate angles but did not otherwise
change the timing or beam pointing direction. We found that the same level of polarization error
was obtained. This also verifies
the procedures used to zero the $B_{Ex}$ and $B_{Ey}$ environmental field components,
since it shows that the bias field does in fact reverse direction after a half period.

\section{Conclusion}

Using the methods discussed above, we have demonstrated control of the magnetic 
field in a TOP trap with 10 mG precision, and we have demonstrated the ability to
apply polarized light to the trapped atoms with errors below $10^{-4}$. These
values are sufficient for our proposed tune-out wavelength experiments, but we
briefly discuss here how much more improvement is possible.

In the case of the rf spectroscopy technique, the sensitivity is fundamentally limited
by the rf pulse duration. To measure frequencies of $2\Omega_1$, the maximum usable 
pulse duration
is a quarter period. At our TOP frequency, this gives a Fourier-limited bandwidth
of 30~kHz. It is reasonable to measure the line center to 1\% of the width, but
beyond that it will likely be necessary to develop a more complex model accounting for
effects like non-uniformity of the bias field, non-idealities of the rf pulse, and
effects of the $B_2$ field. 
At a line-splitting accuracy of 1\%, the 300 Hz frequency resolution corresponds to 
$\delta B = 0.4$~mG. 

The trap position method is limited by the ability to measure the position of the 
Bose condensate. As the dc quadrupole $B_Q'$ is varied, the maximum atom displacement
is $\delta y = 2^{-3/2} B_{Ey}/B_2'$. If $B_2'$ gradient is made too small, then 
it is difficult to ensure the atoms adiabatically follow the trap bottom as $B_Q'$
changes, but a reduction to $B_2' \approx 1$~G/cm is reasonable. At that confinement,
the Thomas--Fermi size of the condensate along $y$ would be 50~$\mu$m, and it is feasible
to measure the condensate position with an accuracy of about 5~$\mu$m. The corresponding
uncertainty in $B_{Ey}$ is then about 1~mG. We conclude that, overall, it should be possible
to reach a performance level for the magnetic field at the mG level without dramatic changes to the measurement techniques presented here. It would be relatively straightforward
to reduce the $B_E$ environmental fields further using magnetic shielding techniques.

Static magnetic traps
can achieve field stabilities of 10~$\mu$G \cite{Dedman07}. This is well below
the variations we achieve, although we expect the time-varying components of our 
field to have comparable amplitude 
stability \cite{Baranowski06}. 
We expect a TOP trap
will be attractive for experiments where the benefits of the rapid field rotation outweigh
the impacts of the corresponding mG-level variations.

The ultimate limit on the polarization accuracy is harder to assess.
One limitation is scattered light from optics and vacuum windows,
which can be challenging to suppress at the $10^{-6}$ level. Another  
limit is set by the spatial uniformity of the retardance. Stress-induced
birefringence is typically non-uniform \cite{schott19}, and retardance variations 
of a few mrad across the laser beam would limit the polarization purity to $10^{-5}$. 
If an optic has a peak retardance of tens of mrad that varies on the cm scale, then 
the variations could be significant for a mm-diameter laser beam. 
To our knowledge, the polarization errors demonstrated here are comparable to what
is typically achieved in a static magnetic trap. 

In summary, we have demonstrated techniques to optimize the magnetic field and light 
polarization in a TOP trap, reaching accuracies of 10 mG and $10^{-4}$ respectively.
These techniques will be important for our own proposed tune-out
wavelength measurements, but we hope that they will also be of use for other
experiments that could benefit from the special features of the TOP trap.

\begin{acknowledgments}

This work was supported by the National Science Foundation (Grant No. PHY-1607571).
We are grateful to Seth Berl for coding support and to Eddie Moan for helpful conversations
and comments on the manuscript.

\end{acknowledgments}

\bibliographystyle{apsrev4-1}
\bibliography{fallon}

\end{document}